\documentclass[conference]{IEEEtran}
\IEEEoverridecommandlockouts
\usepackage{cite}
\usepackage{amsmath,amssymb,amsfonts}
\usepackage{graphicx}
\usepackage{textcomp}
\usepackage{xcolor}
\usepackage{multirow}
\usepackage{subcaption}
\usepackage{algorithm}
\usepackage{algpseudocode}
\def\BibTeX{{\rm B\kern-.05em{\sc i\kern-.025em b}\kern-.08em
    T\kern-.1667em\lower.7ex\hbox{E}\kern-.125emX}}
\begin{document}

\newcommand\Tstrut{\rule{0pt}{2.5ex}}        
\newcommand\Bstrut{\rule[-1ex]{0pt}{0pt}}    
\newcommand\Tstrutt{\rule{0pt}{3ex}}         
\newcommand\Bstrutt{\rule[-1.5ex]{0pt}{0pt}} 

\title{Cross-Environment Transfer Learning for Location-Aided Beam Prediction in 5G and Beyond Millimeter-Wave Networks\\
\thanks{This work has received funding by the German Federal Ministry of Education and Research (BMBF) in the course of the 6GEM research hub under grant number 16KISK038.}}

\author{
	\IEEEauthorblockN{
		Enrico Tosi\IEEEauthorrefmark{2}, 
		Panwei Hu\IEEEauthorrefmark{2}, 
		Aleksandar Ichkov\IEEEauthorrefmark{2},
        Marina Petrova\IEEEauthorrefmark{1}\IEEEauthorrefmark{2},
		Ljiljana Simi\'{c}\IEEEauthorrefmark{2}
	}
	\IEEEauthorblockA{
		\IEEEauthorrefmark{2}
		\textit{Institute for Networked Systems, RWTH Aachen University}
	}
	\IEEEauthorblockA{
		\IEEEauthorrefmark{1}
		\textit{Mobile Communications and Computing, RWTH Aachen University}
	}
	\IEEEauthorblockA{
		\IEEEauthorrefmark{2}
		\{eto, pwu, aic, lsi\}@inets.rwth-aachen.de
        \IEEEauthorrefmark{1}
		petrova@mcc.rwth-aachen.de
	}
	
}

\maketitle

\begin{abstract}
Millimeter-wave (mm-wave) communications require beamforming and consequent precise beam alignment between the gNodeB (gNB) and the user equipment (UE) to overcome high propagation losses. This beam alignment needs to be constantly updated for different UE locations based on beam-sweeping radio frequency measurements, leading to significant beam management overhead. One potential solution involves using machine learning (ML) beam prediction algorithms that leverage UE position information to select the serving beam without the overhead of beam sweeping. However, the highly site-specific nature of mm-wave propagation means that ML models require training from scratch for each scenario, which is inefficient in practice. In this paper, we propose a robust cross-environment transfer learning solution for location-aided beam prediction, whereby the ML model trained on a reference gNB is transferred to a target gNB by fine-tuning with a limited dataset. Extensive simulation results based on ray-tracing in two urban environments show the effectiveness of our solution for both inter- and intra-city model transfer. Our results show that by training the model on a reference gNB and transferring the model by fine-tuning with only 5\% of the target gNB dataset, we can achieve 80\% accuracy in predicting the best beam for the target gNB. Importantly, our approach improves the poor generalization accuracy of transferring the model to new environments without fine-tuning by around 75 percentage points. This demonstrates that transfer learning enables high prediction accuracy while reducing the computational and training dataset collection burden of ML-based beam prediction, making it practical for 5G-and-beyond deployments.  


\end{abstract}


\section{Introduction}

The rapid evolution of wireless communication technologies has led to the exploration of millimeter-wave (mm-wave) spectrum, which provides substantial bandwidth and enables ultra-high data rates and low latency essential for next-generation cellular networks~\cite{8766143}. However, mm-wave frequencies require directional beamforming to overcome high propagation loss. In turn, the sparse nature of the mm-wave channel necessitates precise beam alignment between the gNodeB (gNB) and user equipment (UE). Since the feasible line-of-sight (LoS) and non-LoS (NLoS) propagation paths depend on the urban geometry of the network site, this beam alignment must be constantly updated for different UE locations, resulting in high beam management overhead \cite{NOR2022102947}. 

Machine learning (ML) algorithms have been proposed as a solution for predicting the best beam-pair-link (BPL), thus minimizing the associated overheads (see~\cite{s23094359} and references therein). One promising approach leverages UE location information as training data, capitalizing on the inherent directionality of mm-wave to improve beam prediction accuracy, as explored in~\cite{morais2022positionaidedbeamprediction, 9149272}. However, the ability of these location-aided beam prediction algorithms to generalize across different scenarios has not been investigated. Namely, the prior works on ML-based location-aided beam prediction showed high accuracy when trained and tested on the same gNB, but the generalization accuracy of these models, trained on one gNB but tested on another, is expected to be poor given the site-specific nature of mm-wave propagation. This is a crucial aspect, as training a new ML model from scratch for every scenario is highly inefficient in practice, requires datasets from up to several thousand locations which are costly to acquire, and poses significant computational overhead~\cite{s23094359, morais2022positionaidedbeamprediction}. Transfer learning has emerged as a powerful technique that allows ML models trained for one task to be adapted for another related task through fine-tuning with minimal retraining~\cite{9388790}. While some prior works in~\cite{9580346, 9048929} demonstrated the benefit of transfer learning in enabling fast model readaptation in the context of mm-wave beam management, they focus on transferring knowledge between related tasks within the same gNB. Consequently, they do not explore the transfer capabilities of their algorithms across different physical scenarios, i.e., different gNBs or network environments, leaving open the question of whether ML-based beam management knowledge can be transferred.


To address this important gap, we propose a robust cross-environment transfer learning solution based on a common fully connected neural network (NN) model that inputs UE location information and predicts the best BPL for serving that location. To the best of our knowledge, we are the first to tackle the challenge of transferring the acquired location-aided beam prediction knowledge between gNBs in different physical environments, both within the same city or across cities with varying degrees of urbanization. We consider training the ML model on a reference gNB and then transferring it to a target gNB by fine-tuning the model with a much more limited dataset on the target gNB. We present extensive simulation results, based on ray-tracing in mm-wave networks of 27 gNBs in the real urban environments of Frankfurt and Seoul, to show the excellent transfer learning performance both for inter- and intra-city model transfer. Specifically, our results show that by training the model on a reference gNB and then transferring the model by fine-tuning with only the 5\% of the target gNB dataset, we achieve up to 80\% accuracy in correctly predicting the best BPL for the target gNB. We note that our approach improves the very poor generalization accuracy of simply transferring the model to new environments without fine-tuning by around 75 percentage points. This demonstrates that our proposed transfer learning solution enables high beam prediction accuracy while reducing the computational and training dataset collection burden of ML-based beam prediction, thus making it suitable for practical 5G-and-beyond network deployments.


The rest of the paper is organized as follows. Section~\ref{Sec:System} presents the system model. Section~\ref{Sec:Problem_and_solution} formulates the BPL prediction problem and presents our transfer learning solution. Section~\ref{Sec:results} presents simulation results to evaluate the BPL prediction accuracy. Sec.~\ref{Sec:conclusions} concludes the paper. 

\section{System Model }
\label{Sec:System}

\subsection{Network Model}

\begin{figure}
	\centering
	\begin{subfigure}{0.49\columnwidth}
		\centering
		\includegraphics[width=\linewidth]{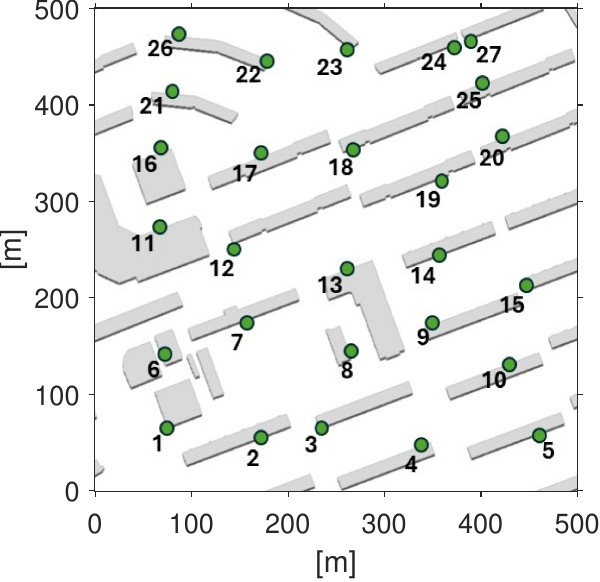}
		\caption{Seoul.}
		\label{fig:SeoulTX}
	\end{subfigure}
	\begin{subfigure}{0.49\columnwidth}
		\centering
		\includegraphics[width=\linewidth]{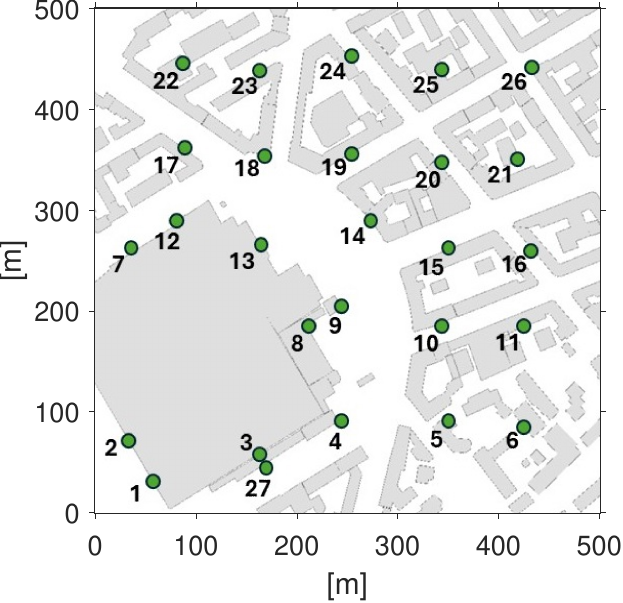}
		\caption{Frankfurt. }
		\label{fig:FrankfurtTX}
	\end{subfigure}
	\caption{Building layout (grey) and gNB locations (green) in our two urban network study areas.}
	\label{fig:beamID}
\end{figure}

We consider an urban downlink mm-wave cellular network consisting of 27 gNBs placed at the building corners at a height of $10$ $\text{m}$, and distributed in a roughly uniform manner in a study area of $500$~$\text{m}$~$\times$~$500$~$\text{m}$. We assume a carrier frequency $f_c=28$~GHz and a nominal gNB transmit power of $P_{\text{tx}}=20$~$\text{dBm}$. We select two study areas with different levels of urbanization: an open space commercial district in Seoul, as shown in Fig.~\ref{fig:SeoulTX}, and a building-dense area around the central station in Frankfurt, as shown in Fig.~\ref{fig:FrankfurtTX}. 

We assume phased antenna arrays of size $8$~$\times$~$8$ at the gNB and $4$~$\times$~$4$ at the UE,  employing codebook-based analog beamforming to cover the complete angular space. We assume a codebook of size $N_{gNB}=64$ at the gNB and $N_{UE}=16$ at the UE, corresponding to a beamwidth of $5.6^{\circ}$ and $22.5^{\circ}$, respectively. We define a directional BPL between a given gNB/UE pair as $l_{i,j}$, where $i$ denotes the gNB beam ID and $j$ denotes the UE beam ID. In total, there are $64$~$\times$~$16$ potential candidate BPLs. We use real 3D antenna patterns obtained via lab measurements with commercially available mm-wave phased antenna arrays~\cite{9918162}.

\subsection{Channel Model}
\label{Sec:ch_model}

 We use the commercial ray-tracing simulator Wireless InSite~\cite{remcom_wireless_insite} to obtain site-specific channel data based on publicly available 3D building models shown in Fig.~\ref{fig:SeoulTX} and Fig.~\ref{fig:FrankfurtTX} for Seoul and Frankfurt, respectively. Channel information is collected with a grid resolution of $1\text{m}$~$\times$~$1\text{m}$ with a ray-launching granularity of $0.1^{\circ}$. Given the dominant propagation characteristics of mm-wave, diffraction and scattering are neglected, while the maximum number of reflections is set to four~\cite{9217146}. We model the buildings using the standardized international telecommunication union (ITU) glass as the material. 

For each considered gNB, the ray-tracing output consists of all the propagation paths from that gNB to each potential UE location within the area. Given a propagation path $k$, the ray-tracer provides: (\textit{i}) the type of path discriminating between LoS and NLoS, (\textit{ii}) angle of arrival (AoA), ~$\{\phi^{\text{UE}}_{ k},\theta^{\text{UE}}_{ k}\}$, angle of departure (AoD), ~$\{\phi^{\text{gNB}}_{ k},\theta^{\text{gNB}}_{ k}\}$ in the azimuth and elevation respectively, and (\textit{iii}) total path loss considering free-space path loss plus any reflection loss. To derive the directional RSS values from this omnidirectional output, we align the antenna patterns of gNB and UE by applying different antenna codebook entries for each steering angle combination. Specifically, the RSS associated with a specific gNB/UE BPL combination $\{i,j\}$ is computed by considering the contribution of all the propagation paths $K$ for an arbitrary UE location, as given by:

\begin{equation}
    \text{RSS}_{i,j} = P_{\text{tx}}\left(\sum_{k=1}^{K}PL_{k}G^{\text{gNB}}_{i}(\phi^{\text{gNB}}_{k}, \theta^{\text{gNB}}_k)G^{\text{UE}}_{j}(\phi^{\text{UE}}_{k}, \theta^{\text{UE}}_k)\right)
	\label{eq:RSS}
\end{equation}

\noindent where $PL_k$ denotes the total path loss for path $k$, $G^{\text{gNB}}_{i}(\phi^{\text{gNB}}_{k}, \theta^{\text{gNB}}_k)$ is the gNB antenna gain for the $i$-th codebook beam ID in the direction $\{\phi^{\text{gNB}}_{k}, \theta^{\text{gNB}}_k\} $ corresponding to the AoD of propagation path $k$, and $G^{\text{UE}}_{j}(\phi^{\text{UE}}_{k}, \theta^{\text{UE}}_k)$ is the UE antenna gain for the $j$-th codebook ID in the direction $\{\phi^{\text{UE}}_{k}, \theta^{\text{UE}}_k\}$ corresponding to the AoA of the same propagation path. 

Using (\ref{eq:RSS}), we build the BPL matrix of size $N_{gNB} \times N_{UE}$ containing the RSS value for every candidate BPL $l_{i,j}$. We extract the top-5 best BPLs from the BPL matrix, i.e., the gNB/UE codebook entry pairs corresponding to the five BPLs having the highest RSS values. Each top-5 set, tagged with the corresponding UE location in the network area covered by an arbitrary $\text{gNB}_x$, constitutes the BPL dataset $D_{gNB_{x}}$ for that gNB. We note that we consider locations to be covered only if the maximum RSS value in the corresponding BPL matrix is greater than a minimum RSS threshold ($\text{RSS}_{thresh} = -174$ dBm); otherwise, they are considered to be locations in outage and excluded from the dataset $D_{gNB_{x}}$. Consequently, the sizes of these datasets vary for different gNBs, as shown in Fig.~\ref{fig:Frankfurt_LoS} and Fig.~\ref{fig:Seoul_LoS} for Frankfurt and Seoul gNBs, respectively.

\section{Beam Prediction Problem Formulation \& Proposed Transfer Learning Solution}
\label{Sec:Problem_and_solution}

\subsection{Beam Prediction Problem Formulation}

In our communication system, both the gNB and the UE have beamsteering capability using predefined codebooks. Let $i$ be the codebook entry at the gNB, selected from the codebook $\mathcal{F} = \{i_m\}_{m=1}^{N_{gNB}}$, and $j$ be the codebook entry at the UE, from the codebook $\mathcal{W} = \{j_n\}_{n=1}^{N_{UE}}$. The beam selection aims to identify the optimal BPL $l_{i^*,j^*}$ consisting of the gNB and UE codebook entry pair $(i^*,j^*)$ that maximizes the RSS in (\ref{eq:RSS}). However, rather than explicitly calculating the optimal beam codebook pair $(i^*,j^*)$ using exhaustive beam sweeping measurements as per FR2 initial access procedure~\cite{NOR2022102947} which incur in a high overhead, we aim to estimate the beam codebook pair $(\hat{i},\hat{j})$ that maximizes the probability of selecting the optimal codebook pair based solely on the UE position $\{x,y\}$:

\begin{equation}
    (\hat{i}, \hat{j}) = \arg \max_{i \in \mathcal{F}, j \in \mathcal{W}} P((i, j) = (i^*,j^*) \ | \{x,y\})
\end{equation}

\subsection{Proposed Location-Based Transfer Learning Solution}
\label{Sec:solution}

\begin{algorithm}[b]
    \caption{Cross-Environment Transfer Learning}
    \label{alg:TL}
    \begin{algorithmic}[1]
        \Require $D_{gNB_x}$: BPL dataset for $\text{gNB}_x$
        
        $D_{gNB_y}$: BPL dataset for $\text{gNB}_y$
        \Ensure{$w_{gNB_y}$: Model parameters for $\text{gNB}_y$}

        \State \textbf{Step1: Model Training for $\text{gNB}_x$} 
        \State Training, validation and testing split of $D_{gNB_x}$
        \State Model training using training set of $D_{gNB_x}$ 
        \State $w_{gNB_x} \leftarrow$ Save model parameters
        \State \textbf{Step 2: Model Transfer to $\text{gNB}_y$}
        \State Initialize the model with parameters $w_{gNB_x}$
        \State $\Tilde{D}_{gNB_y} \leftarrow$ subsampling $D_{gNB_y}$
        \State Training, validation and testing split of $\Tilde{D}_{gNB_y}$ 
        \State Fine-tuning using training set of $\Tilde{D}_{gNB_y}$
        \State $w_{gNB_y} \leftarrow$ Save fine-tuned model parameters for $\text{gNB}_y$
    \end{algorithmic}
\end{algorithm}

We propose to learn the mapping function between the UE location and the best BPL serving that location for a reference $\text{gNB}_x$ using a standard fully-connected NN and transfer the knowledge acquired by the model to another target $\text{gNB}_y$, as described in Alg.~\ref{alg:TL}. In the first step, we consider the complete BPL dataset $D_{gNB_x}$ specific to the reference $\text{gNB}_x$ and we split it into training, validation, and testing sets with a ratio of $[60\%, 20\%, 20\%]$. We train the NN using the training set and save the resulting weights $w_{gNB_x}$. In the second step, the NN initialized with $w_{gNB_x}$ is transferred to the target $\text{gNB}_y$ and we now consider a reduced dataset $\Tilde{D}_{gNB_y}$ uniformly sampled from the corresponding BPL dataset $D_{gNB_y}$. We split this reduced dataset into training, validation, and testing with the same ratio, and we use the training set for fine-tuning the model. Finally, the fine-tuned weights  $w_{gNB_y}$ are saved, enabling $\text{gNB}_y$ to predict the BPL for a new UE once its location is provided.

We select a fully-connected NN as in~\cite{morais2022positionaidedbeamprediction}, whose structure and hyper-parameters are reported in Table \ref{tab:NNstructure}. In~\cite{morais2022positionaidedbeamprediction}, the UE location is used as a training feature to select the best gNB codebook entry in a search space of size $N_{gNB}$, while assuming an omni-directional UE antenna. In contrast, we assume directional antennas at both the gNB and UE sides, resulting in a search space of size $N_{gNB} \times N_{UE}$. Given the more challenging beam selection task, we evaluate the accuracy of correctly predicting the best BPL, i.e., top-1 accuracy, as well as the accuracy of the best BPL being within the top-5 BPLs predicted by the model, i.e. best-in-top-5 accuracy. We emphasize that the selected best-in-top-5 metric allows us to evaluate the possibility of narrowing the search space to only five BPLs instead of performing a full sweep over the complete search space. Therefore, our training labels comprise the top five BPLs for the given UE location.

To address the limitation of the cross-entropy loss function, commonly used for classification tasks and which only favors the best prediction, we employ the weighted cross-entropy loss function given by

\begin{equation}
    L = \sum_{k=1}^5 -w_k \log \hat{p}_k
\end{equation}

\noindent where $\hat{p}_k$ denotes the predicted probability that the $\text{k}^{th}$ best BPL is selected, and $w_k$ is the corresponding weight. This approach prioritizes the best BPL selection while also promoting the selection of one subsequent top-5 BPL through the weighted scheme $[0.5, 0.2, 0.15, 0.075, 0.075]$.

We evaluate the effectiveness of our cross-environment transfer learning by examining the top-1 and best-in-top-5 accuracies on the testing set relative to the target gNB. 

\begin{table}
	\caption{NN architecture and hyperparameters}
    \centering

\begin{tabular}{ | c | c |  }

 \hline
 
 \textbf{Parameters} & \textbf{Values} \\
 \hline
 Input size   & $2$ $\{x, y\}$   \\
 Label size   & 5                           \\
 Hidden Layers &  3 layers, 128 nodes each  \\
 Output size & $64 \times 16$ (number of available BPLs)  \\
 Activation    & ReLu on hidden layers\\
 Loss function   & Customized weighted cross-entropy loss \\
 Training/Validation/Test & $[60\%, 20\%, 20\%]$ \\
 Training batch size &   128  \\
 Learning epochs & 60 \\ 
 Optimizer & Adam \\
 Initial learning rate & 0.2 \\
 Learning rate scheduler & MultiStepLR, steps at 20 and 40 \\
 \hline
\end{tabular}
\label{tab:NNstructure}
\vspace{-0.5cm}
\end{table}

\section{Results}
\label{Sec:results}
In this section we present a performance evaluation of our transfer learning solution proposed in Sec.~\ref{Sec:Problem_and_solution}. Sec.~\ref{Sec:intra} presents the performance of transfer learning from a reference gNB to a target gNB in the same city, i.e., intra-city transfer. Sec.~\ref{Sec:cross} extends the analysis by evaluating the performance of transferring between gNBs in different cities, i.e., inter-city transfer. In Sec.~\ref{Sec:fineTuning} we study how the amount of fine-tuning data impacts the performance.


\subsection{Intra-City Transfer Learning Performance}
\label{Sec:intra}

\begin{figure}[t]
	\centering
	\begin{subfigure}{0.45\columnwidth}
		\centering
		\includegraphics[width=\linewidth]{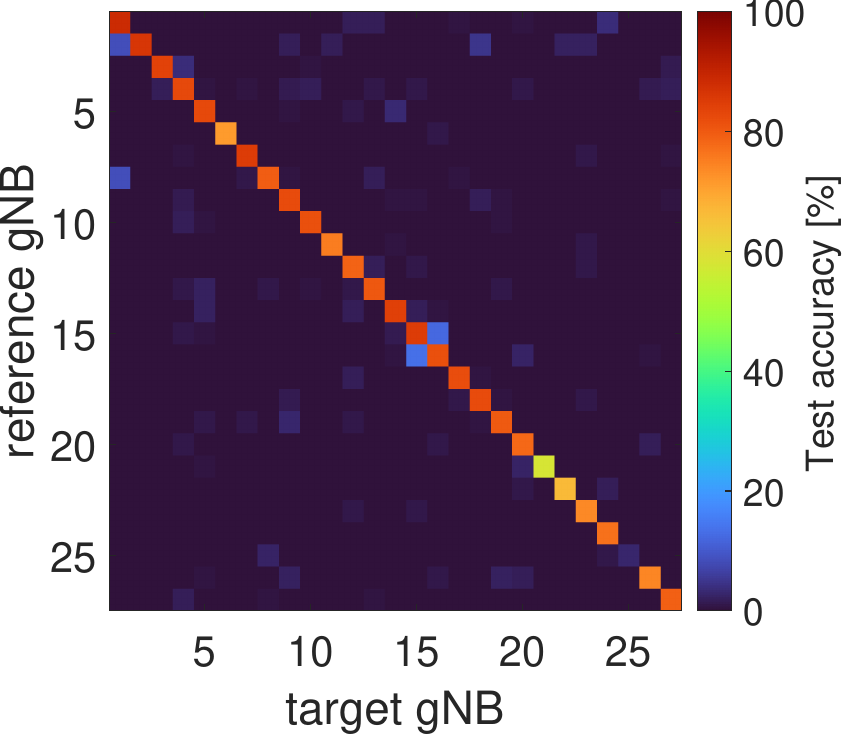}
		\caption{Frankfurt generalization.}
		\label{fig:FrankGen}
	\end{subfigure}
	\hspace{0.5cm}
        \begin{subfigure}{0.45\columnwidth}
		\centering
		\includegraphics[width=\linewidth]{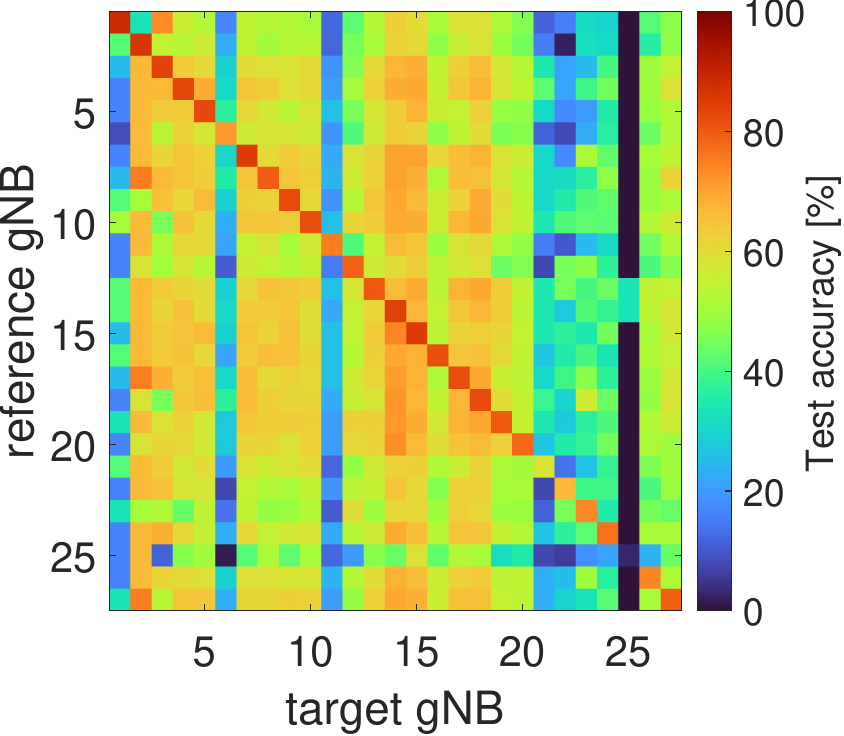}
		\caption{Frankfurt transfer.}
		\label{fig:Frank_confusion}
	\end{subfigure}
	
    \hfill
	\begin{subfigure}{0.45\columnwidth}
		\centering
		\includegraphics[width=\linewidth]{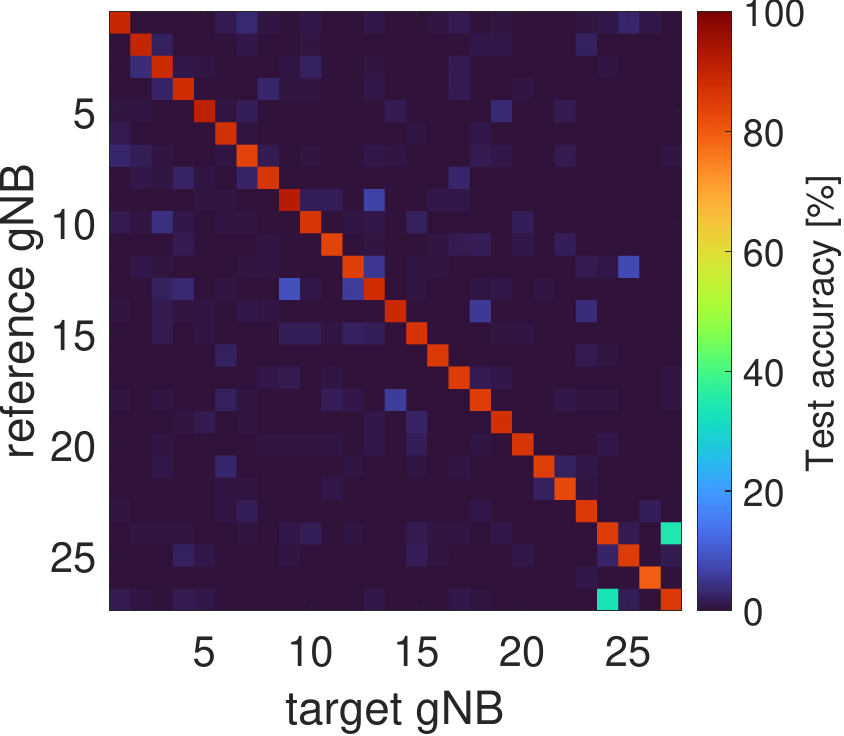}
		\caption{Seoul generalization.}
		\label{fig:SeoulGen}
	\end{subfigure}
        \hspace{0.5cm}
	\begin{subfigure}{0.45\columnwidth}
		\centering
		\includegraphics[width=\linewidth]{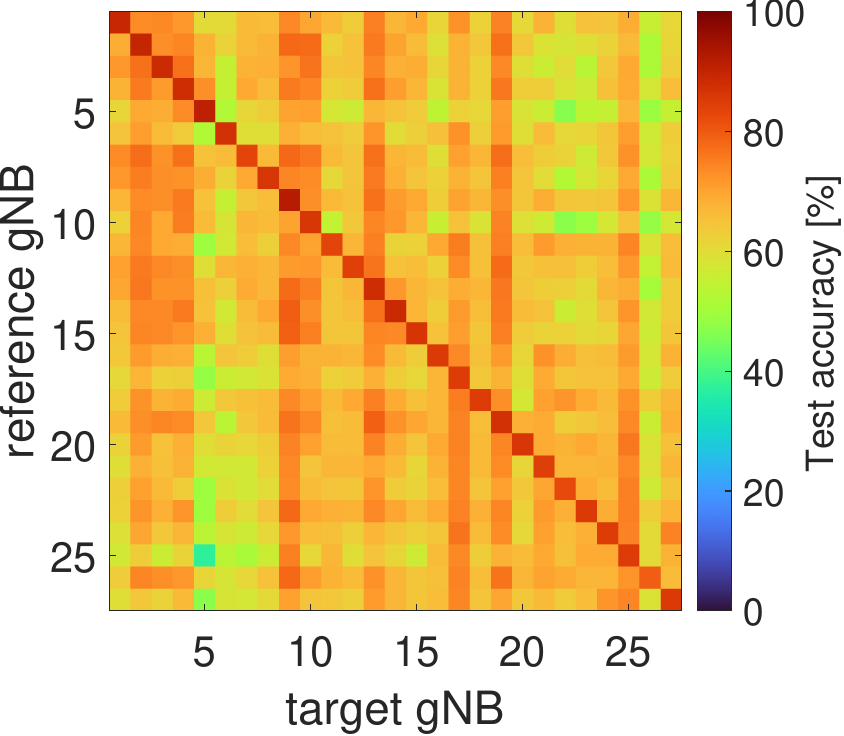}
		\caption{Seoul transfer.}
		\label{fig:Seoul_confusion}
	\end{subfigure}
	\caption{Top-1 BPL prediction: generalization and transfer learning accuracy using 5\% fine-tuning dataset size across all gNB combinations of reference and target gNBs in (a)-(b) Frankfurt and (c)-(d) Seoul. The diagonal entries correspond to the accuracy of training and testing on the same gNB.  }
    \vspace{-0.35cm}
	\label{fig:confusion}
\end{figure}



\begin{figure}[t]

	\centering
	\begin{subfigure}{0.98\columnwidth}
		\centering
		\includegraphics[width=\linewidth]{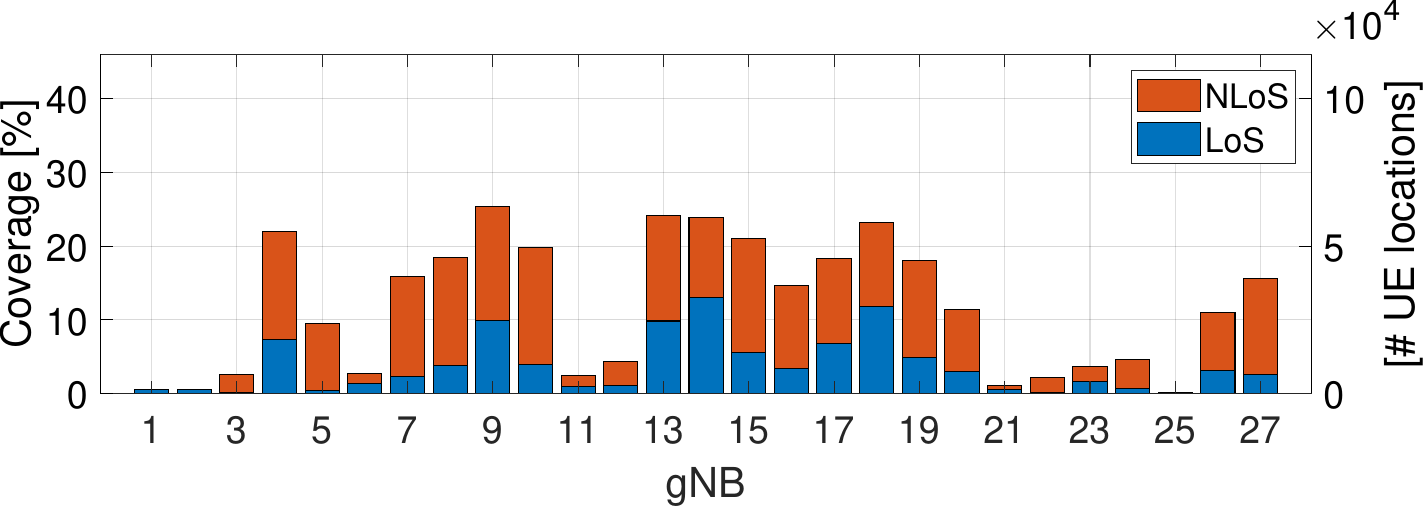}
		\caption{Frankfurt.}
		\label{fig:Frankfurt_LoS}
	\end{subfigure}
	\hfill
	\begin{subfigure}{0.98\columnwidth}
		\centering
		\includegraphics[width=\linewidth]{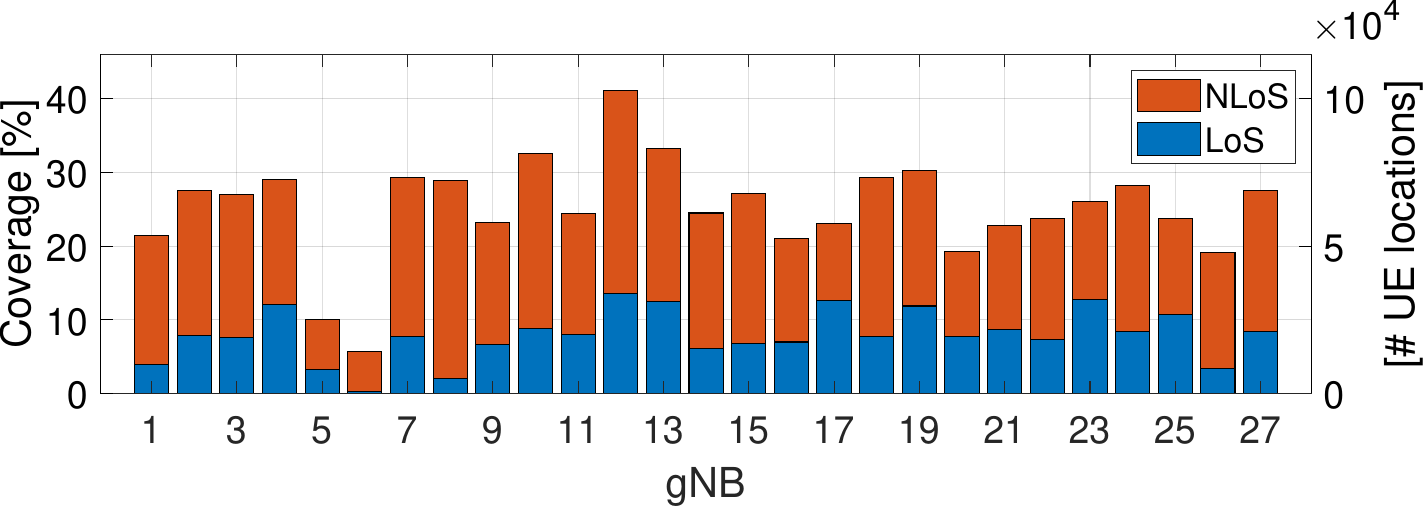}
		\caption{Seoul.}
		\label{fig:Seoul_LoS}
	\end{subfigure}
	\caption{Percentage coverage with respect to the total $500$~$\text{m}$~$\times$~$500$~$\text{m}$ study area and corresponding dataset size of the gNBs in (a) Frankfurt and (b) Seoul, differentiating between LoS and NLoS coverage.}
    \vspace{-0.35cm}
	\label{fig:LoS}
\end{figure}

Let us first evaluate the generalization accuracy of our model by training it for a reference gNB and testing it on a target gNB \textit{without} fine-tuning. Fig.~\ref{fig:FrankGen} shows this result in terms of top-1 BPL prediction accuracy for all the gNB combinations in Frankfurt. We note that the diagonal entries correspond to the baseline accuracy of the model, i.e., training and testing on the same gNB. Fig.~\ref{fig:FrankGen} shows that the model generalizes very poorly, with a mean top-1 accuracy of 5\% for gNB combinations where the reference and target gNBs do not coincide. This clearly demonstrates the site-specific nature of the learning task and highlights the necessity of fine-tuning data specific to the target gNB for effective transfer learning. Fig.~\ref{fig:Frank_confusion} shows the top-1 accuracy achieved by transferring the model between gNB combinations in  Frankfurt and fine-tuning it with only 5\% of the target gNB dataset. Again, the diagonal entries of the matrix represent the baseline accuracy achieved by training and testing on the same gNB. Comparing Fig~\ref{fig:FrankGen} and Fig.~\ref{fig:Frank_confusion} shows that transfer learning enables the model to adapt better to the new environment, significantly improving the top-1 beam prediction accuracy. Specifically, the transfer learning accuracy is up to 75\%, corresponding in an improvement of 70 percentage points from the generalization accuracy in Fig.~\ref{fig:FrankGen}. Fig.~\ref{fig:Frank_confusion} also reveals a distinct vertical line pattern for most target gNBs, indicating that the model adapts better for some target gNBs than others. To investigate the underlying reasons, let us consider Fig.~\ref{fig:Frankfurt_LoS}, which shows the coverage characteristics for the gNBs in Frankfurt. We observe that the gNBs covering smaller areas in Fig.~\ref{fig:Frankfurt_LoS} generally correspond to the target gNBs with low transfer learning accuracy in Fig.~\ref{fig:Frank_confusion}, i.e., gNBs 1, 6, 11, 20, 21, and especially 25. This is because less extensive coverage corresponds to a reduced dataset size and likely insufficient data for fine-tuning the model. This is also consistent with the lower baseline accuracy (diagonal in Fig.~\ref{fig:FrankGen} and Fig.~\ref{fig:Frank_confusion}) for these gNBs. Fig.~\ref{fig:Frank_confusion} also shows that this group of gNBs with small coverage generally perform worse as the reference gNB, suggesting that the size of the dataset plays a key role also in selecting which reference gNB the model should be transferred from. 

We conduct the same analysis in Seoul to validate the aforementioned insights, with Fig.~\ref{fig:SeoulGen} and Fig.~\ref{fig:Seoul_confusion} presenting the generalization and transfer accuracies, respectively. Fig.~\ref{fig:SeoulGen} confirms that the model exhibits very poor generalization also in Seoul, as evidenced by the low top-1 accuracies outside the diagonal. Fig.~\ref{fig:Seoul_confusion} shows that transfer learning with 5\% fine-tuning leads to substantial accuracy improvements in Seoul, resulting in beam prediction accuracy ranging between 60\% and 80\%, corresponding to an improvement between 55 and 75 percentage points respect to the mean generalization accuracy in Fig.~\ref{fig:SeoulGen}. The transfer learning accuracies in Seoul in Fig.~\ref{fig:Seoul_confusion} often surpass those observed in Frankfurt in Fig~\ref{fig:Frank_confusion}, likely because Seoul gNBs generally cover more extensive areas (\textit{cf}. Fig.~\ref{fig:Frankfurt_LoS} vs Fig.~\ref{fig:Seoul_LoS}), resulting in larger datasets used for training and fine-tuning the model. Furthermore, the accuracies in Seoul in Fig.~\ref{fig:Seoul_confusion} are more homogeneous among the gNB combinations with respect to Frankfurt in Fig.~\ref{fig:Frank_confusion}. However, we still observe that the model achieves lower accuracies when adapting to target gNBs 5, 6 and 26. This is consistent with the small dataset size of gNBs 5 and 6 in Fig.~\ref{fig:Seoul_LoS}, while gNB 26's coverage is comparable to e.g. gNBs 16 and 20 which exhibits better performance in Fig.~\ref{fig:Seoul_confusion}. This suggests that other factors, such as e.g., the spatial distribution of the type of coverage beam orientation, likely play a role in the transfer learning accuracy, although their effect is not straightforward to explain, as evidenced by the lack of clear correlation between the e.g. LoS/NLoS per gNB coverage reported in Fig.~\ref{fig:LoS} and the accuracies in Fig.~\ref{fig:confusion}. Instead, in Sec.~\ref{Sec:fineTuning} we further investigate the impact of the dominant observable factor of the training dataset size.

\subsection{Inter-City Transfer Learning Performance}
\label{Sec:cross}



\begin{figure}[t]
	\centering
	\begin{subfigure}{0.45\columnwidth}
		\centering
		\includegraphics[width=\linewidth]{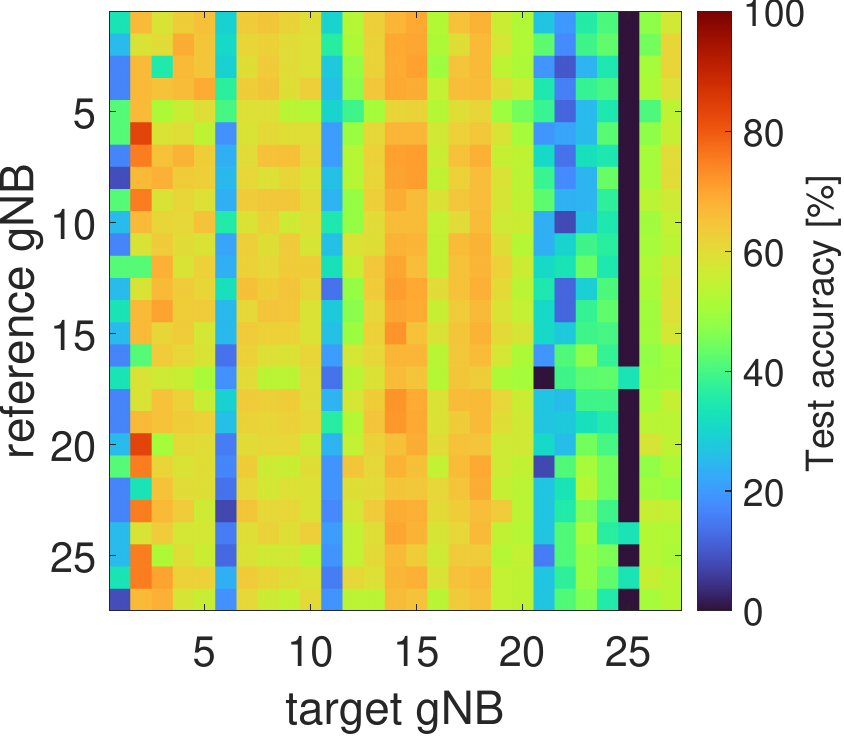}
		\caption{Seoul-to-Frankfurt.}
		\label{fig:Seoul_to_Frank_confusion}
	\end{subfigure}
	\hfill
	\begin{subfigure}{0.45\columnwidth}
		\centering
		\includegraphics[width=\linewidth]{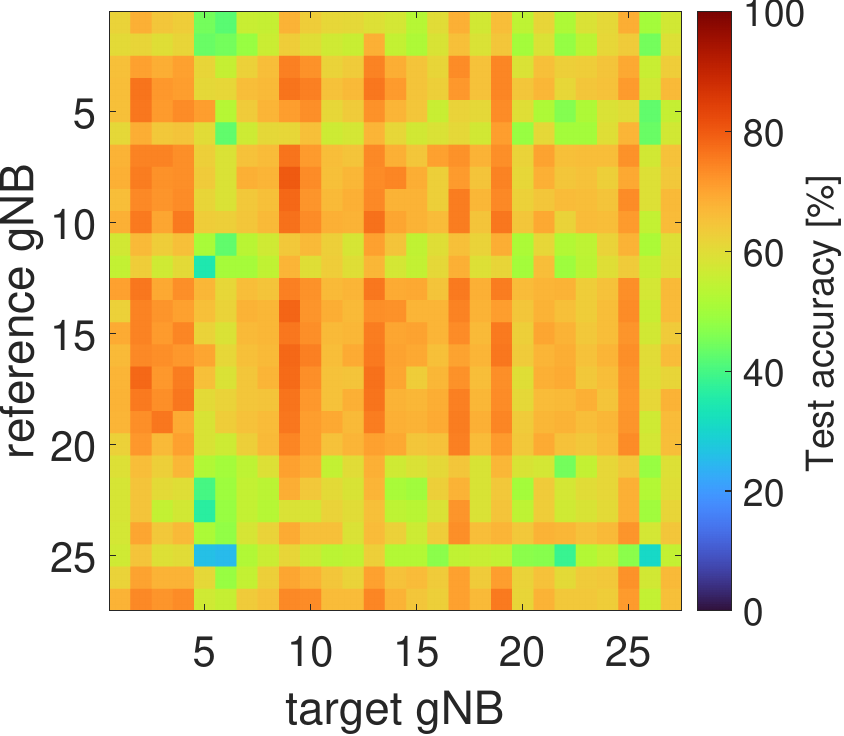}
		\caption{Frankfurt-to-Seoul.}
		\label{fig:Frank_to_Seoul_confusion}
	\end{subfigure}
 \caption{Top-1  transfer learning accuracy across the cities using 5\% fine-tuning for gNB combinations of (a) reference gNBs in Seoul and target gNBs in Frankfurt and (b) vice-versa.  }
 \vspace{-0.4cm}
	\label{fig:confusion_across}
\end{figure}

Let us now examine how transfer learning performs between different cities to determine to what extent our approach be applied between network environments with varying urban densities and layouts. Fig.~\ref{fig:Seoul_to_Frank_confusion} shows the top-1 accuracies achieved when transferring from reference gNBs in Seoul to target gNBs in Frankfurt with a fine-tuning percentage of 5\%, while Fig.~\ref{fig:Frank_to_Seoul_confusion} shows the reverse scenario, i.e. transferring from reference gNBs in Frankfurt to target gNBs in Seoul with the same fine-tuning percentage. Fig.~\ref{fig:Seoul_to_Frank_confusion} shows that the per-target gNB transfer learning accuracy when transferring models trained on Seoul gNBs is comparable to the inter-city accuracy in Frankfurt in Fig.~\ref{fig:Frank_confusion}. The dominant trend in Fig.~\ref{fig:Seoul_to_Frank_confusion} is again the difficulty of adapting to target gNBs 1, 6, 11, 21, 22, and 25, for which the fine-tuning dataset might be insufficient. This suggests that despite the reference gNBs being in another city, the model can adapt well as long as it is provided with sufficient fine-tuning data from the target gNB. Fig.~\ref{fig:Frank_to_Seoul_confusion} shows that gNBs providing small area coverage in Frankfurt, e.g. gNBs 1, 2, 21, 22, and 25 in Fig.~\ref{fig:Frankfurt_LoS}, perform relatively poorly as reference gNBs for transfer learning to Seoul. In contrast, effective transfer learning is enabled by choosing reference gNBs with larger area coverage in Frankfurt, e.g., 10 and 13 in Fig.\ref{fig:Frankfurt_LoS}. Overall, the high inter-city transfer learning accuracy in Fig.~\ref{fig:confusion_across} of up to 80\% is highly encouraging, suggesting that our approach is practically scalable also across different cities.

\subsection{Fine-Tuning Effect}
\label{Sec:fineTuning}


\begin{figure}[t]
	\centering
	\begin{subfigure}{0.98\columnwidth}
		\centering
		\includegraphics[width=\linewidth]{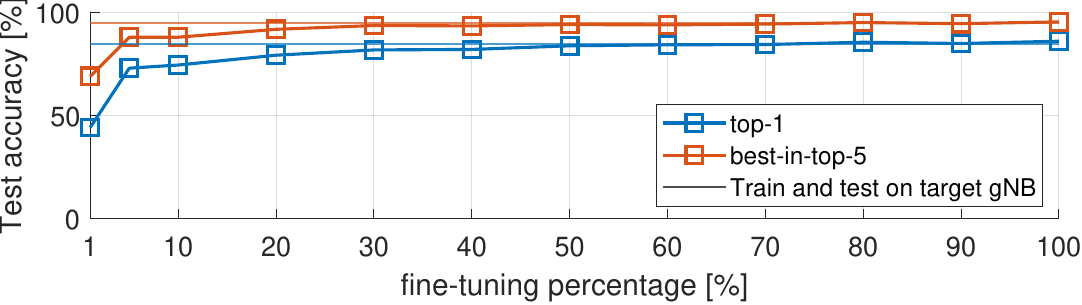}
		\caption{From reference gNB 20 to target gNB 14.}
		\label{fig:Frank_good_incremental}
	\end{subfigure}
	\hfill
	\begin{subfigure}{0.98\columnwidth}
		\centering
		\includegraphics[width=\linewidth]{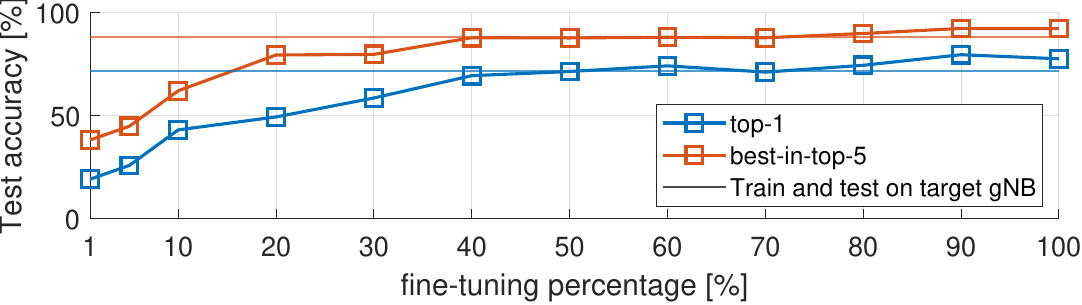}
		\caption{From reference gNB 17 to target gNB 11.}
		\label{fig:Frank_bad_incremental}
	\end{subfigure}
	\caption{Test accuracy vs. percentage of target gNB dataset used for fine-tuning in transferring between selected gNB combinations in Frankfurt. The plots also include the baseline accuracy of training and testing on target gNB.}
	\vspace{-.75cm}
    \label{fig:incremental}
\end{figure}

Let us now explicitly explore the impact of varying the fine-tuning dataset percentage on the transfer learning performance. We focus on the Frankfurt environment given the greater variance of the results among the target gNBs observed in Fig.~\ref{fig:Frank_confusion}. Fig.~\ref{fig:Frank_good_incremental} presents the top-1 transfer accuracy versus the size of fine-tuning dataset as in percentage to the target gNB dataset, for transferring from gNB 20 to gNB 14 in Frankfurt. Similarly, Fig.~\ref{fig:Frank_bad_incremental} presents these results for transferring from gNB 17 to gNB 11 in the same city. Moreover, to fully evaluate the performance of transfer learning in practice, in this section we additionally consider the best-in-top-5 transfer accuracy, as introduced in Sec.~\ref{Sec:solution}. This corresponds to the accuracy of having the best BPL within the top-5 predicted BPLs, allowing us to evaluate the possibility of narrowing the search space to the top-5 BPLs predicted by the fine-tuned model. The corresponding top-1 and best-in-top-5 accuracies for training and testing on the target gNB are reported as a baseline. We emphasize that, as shown in Fig~\ref{fig:Frankfurt_LoS}, gNB 11 has a smaller coverage area than gNB 14 and consequently a smaller overall dataset size (6150 and 59778 respectively). Fig.~\ref{fig:Frank_good_incremental} shows that for target gNB 14 the model reaches top-1 transfer accuracy of 73\% at 5\% fine-tuning and only slightly improves to 80\% with larger fine-tuning percentages, suggesting there is no significant benefit from additional fine-tuning data. The corresponding best-in-top-5 curve follows a similar trend, achieving an excellent accuracy of 88\% at 5\% fine-tuning (vs. 94\% baseline accuracy). By contrast, Fig.~\ref{fig:Frank_bad_incremental} shows that for target gNB 11 the model significantly improves its top-1 transfer accuracy by increasing the fine-tuning percentages beyond 5\%, reaching the corresponding baseline of 71\% at 50\% fine-tuning dataset size and stabilizing around 75\% for higher percentages\footnote{We note that the transfer model can even outperform the baseline as it leverages the complete reference gNB training data plus target gNB fine-tuning, whereas the baseline model only uses data from the target gNB.}. These results confirm that the low transfer learning accuracy in Fig.~\ref{fig:Frank_confusion} for this gNB is due to insufficient fine-tuning data rather than model saturation, as gNB 11's dataset is significantly smaller than those of other gNBs (\textit{cf}. Fig.~\ref{fig:Frankfurt_LoS}). Thus, increasing the fine-tuning percentage for gNBs with smaller coverage areas and datasets should be a means to achieve good transfer learning performance in practice.

\begin{figure}[t]
	\centering
        \begin{subfigure}{0.45\columnwidth}
		\centering
		\includegraphics[width=\linewidth]{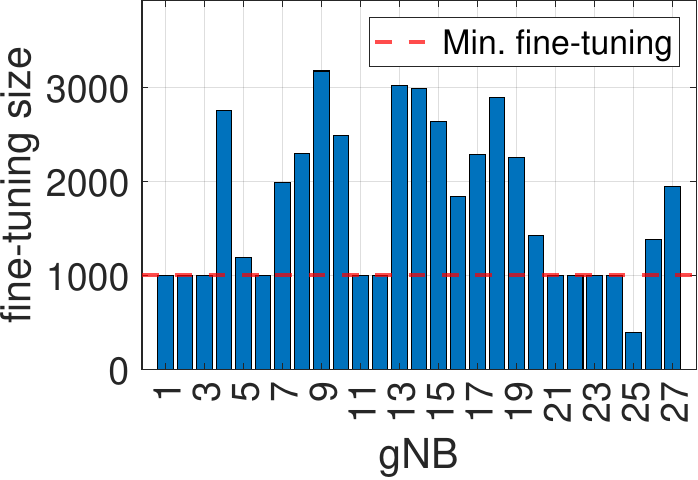}
		\caption{}
		\label{fig:FT_size}
	\end{subfigure}
        \hspace{0.5cm}
	\begin{subfigure}{0.45\columnwidth}
		\centering
		\includegraphics[width=\linewidth]{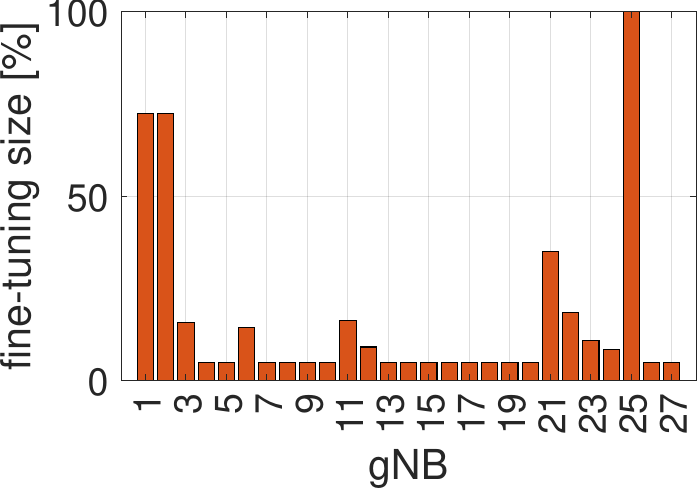}
		\caption{}
		\label{fig:FT_perc}
	\end{subfigure}
 
    \hspace{2pt}
	\begin{subfigure}{0.45\columnwidth}
		\centering
		\includegraphics[width=\linewidth]{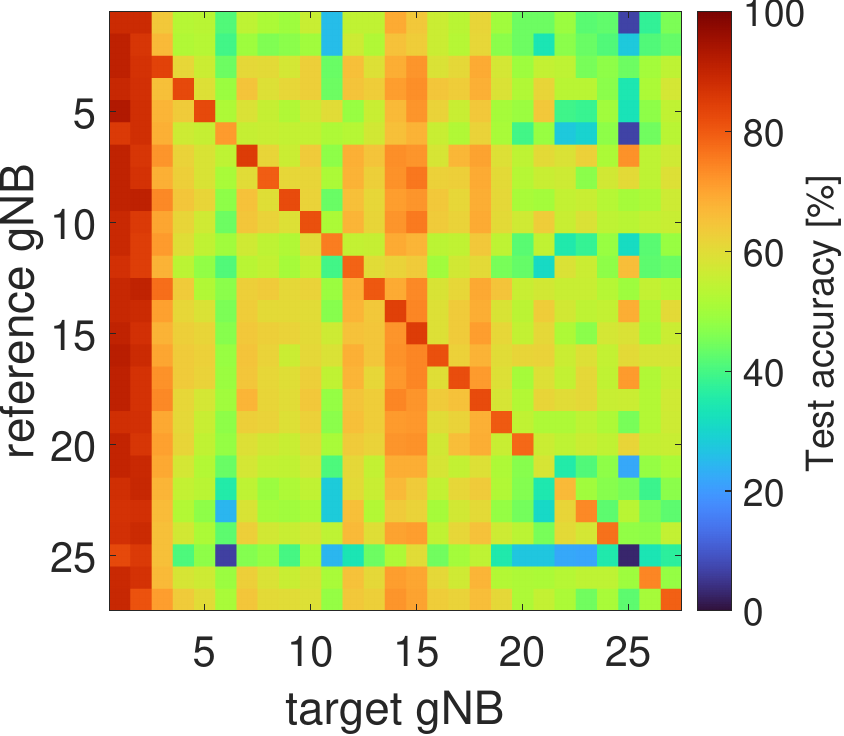}
		\caption{Top-1.}
		\label{fig:Frank_confusion_truncated_top1}
	\end{subfigure}
	\hspace{0.4cm}
        \begin{subfigure}{0.45\columnwidth}
		\centering
		\includegraphics[width=\linewidth]{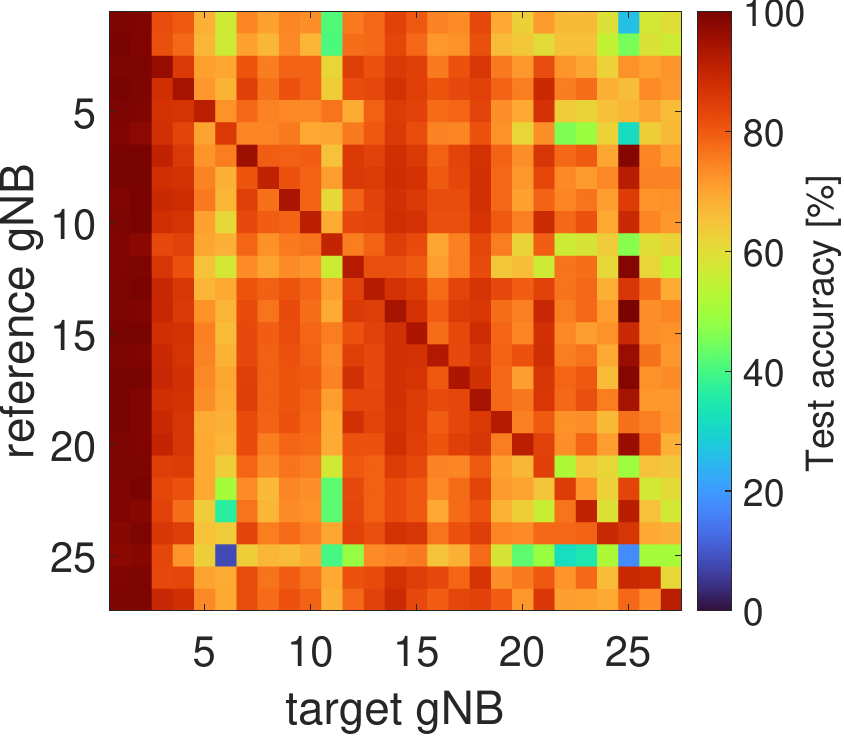}
		\caption{Best-in-top-5.}
		\label{fig:Frank_confusion_truncated_top5}
	\end{subfigure}
	
    \hfill
	\vspace{-0.5cm}
	\caption{(c) Top-1 and (d) best-in-top-5 transfer learning accuracy in Frankfurt with gNB-specific fine-tuning dataset size, given in (a) as absolute dataset size and in (b) as a percentage of to the total dataset size of the target gNB.   }
	\label{fig:confusion_truncated}
    \vspace{-0.35cm}
\end{figure}

To study this, we now consider the performance of our transfer learning solution with a minimum fine-tuning dataset size of 1000. Namely, we set the fine-tuning dataset size to 1000 for all the datasets where 5\% of the total is lower than 1000 and to 5\% otherwise (gNB 25 being the exception, as its dataset consists of only 389 data points). Fig~\ref{fig:FT_size} and Fig.~\ref{fig:FT_perc} show the resulting size of fine-tuning datasets, as absolute size and as a percentage relative to the total dataset size for each gNB (\textit{cf}. Fig.~\ref{fig:Frankfurt_LoS}). Fig.~\ref{fig:Frank_confusion_truncated_top1} and Fig.~\ref{fig:Frank_confusion_truncated_top5} present the resulting top-1 and best-in-top-5 beam prediction transfer learning accuracies for all gNB combinations in Frankfurt. Fig.~\ref{fig:Frank_confusion_truncated_top1} confirms that the target gNBs suffering from lower accuracy in Fig.~\ref{fig:Frank_confusion}, i.e., gNBs 1, 6, 11, 21, 22, and 25, can now better adapt from other models thanks to the additional fine-tuning data. Notably, the very low top-1 baseline accuracy of gNB 25 of 3\% is significantly improved by first transferring the model from any other gNB and then fine-tuning it: e.g. transferring from gNB 7 achieves a top-1 accuracy of 73\%. This suggests that transferring previously trained models on other gNBs can be extremely helpful for gNBs with limited coverage area, provided a reasonable minimum amount of site-specific fine-tuning data. Fig.~\ref{fig:Frank_confusion_truncated_top1} also shows that increasing the fine-tuning for target gNBs 1 and 2 results in very high top-1 accuracies of around 90\%. This may be due to their area coverage being completely LoS (\textit{cf}. Fig.~\ref{fig:Frankfurt_LoS}), making their optimal BPL patterns geometrically uniform and thus easier to learn and adapt to. Finally, Fig.~\ref{fig:Frank_confusion_truncated_top5} shows that the best BPL is generally included in the top five BPLs predicted with very high reliability. In particular, for every target gNB, there exists a reference gNB ensuring best-in-top-5 transfer accuracy above 75\%. Overall, Fig.~\ref{fig:confusion_truncated} demonstrates that our transfer learning approach performs well even in highly site-specific scenarios in densely urbanized areas like Frankfurt, provided there is a sufficient fine-tuning data size which is typically between 5\% and 40\% of the available dataset.

\section{Conclusions}
\label{Sec:conclusions}

We investigated cross-environment transfer learning to enhance BPL prediction efficiency for 5G and beyond mm-wave networks. We proposed training the model on a reference gNB and then transferring it to a target gNB in the same or a different city by fine-tuning the model with a much more limited dataset from the target gNB. Our results show that minimal fine-tuning (5\% of the target gNB dataset) can be sufficient to adapt the model to a new environment and achieve an accuracy of up to 80\% in predicting the best BPL. Moreover, the high accuracy obtained in inter-city transfer between Frankfurt and Seoul suggests that transfer learning is practically scalable also across different cities with varying urban densities and layouts. Finally, we showed that our method remains effective even in highly site-specific scenarios in densely urbanized areas like Frankfurt, provided a reasonable fine-tuning data size is used. Our ongoing work aims to further optimize transfer learning by exploring different learning algorithms and fine-tuning strategies.


\bibliographystyle{IEEEtran}
\bibliography{Bibliography.bib}

\begin{thebibliography}{10}
\providecommand{\url}[1]{#1}
\csname url@samestyle\endcsname
\providecommand{\newblock}{\relax}
\providecommand{\bibinfo}[2]{#2}
\providecommand{\BIBentrySTDinterwordspacing}{\spaceskip=0pt\relax}
\providecommand{\BIBentryALTinterwordstretchfactor}{4}
\providecommand{\BIBentryALTinterwordspacing}{\spaceskip=\fontdimen2\font plus
\BIBentryALTinterwordstretchfactor\fontdimen3\font minus \fontdimen4\font\relax}
\providecommand{\BIBforeignlanguage}[2]{{%
\expandafter\ifx\csname l@#1\endcsname\relax
\typeout{** WARNING: IEEEtran.bst: No hyphenation pattern has been}%
\typeout{** loaded for the language `#1'. Using the pattern for}%
\typeout{** the default language instead.}%
\else
\language=\csname l@#1\endcsname
\fi
#2}}
\providecommand{\BIBdecl}{\relax}
\BIBdecl

\bibitem{8766143}
Z.~Zhang \emph{et~al.}, ``{6G} wireless networks: Vision, requirements, architecture, and key technologies,'' \emph{IEEE Vehicular Technology Magazine}, vol.~14, no.~3, pp. 28--41, 2019.

\bibitem{NOR2022102947}
A.~M. Nor, S.~Halunga, and O.~Fratu, ``Survey on positioning information assisted mmwave beamforming training,'' \emph{Ad Hoc Networks}, vol. 135, p. 102947, 2022.

\bibitem{s23094359}
D.~d.~S. Brilhante \emph{et~al.}, ``A literature survey on ai-aided beamforming and beam management for {5G} and {6G} systems,'' \emph{Sensors}, vol.~23, no.~9, 2023.

\bibitem{morais2022positionaidedbeamprediction}
\BIBentryALTinterwordspacing
J.~Morais, A.~Behboodi, H.~Pezeshki, and A.~Alkhateeb, ``Position aided beam prediction in the real world: How useful gps locations actually are?'' 2022. [Online]. Available: \url{https://arxiv.org/abs/2205.09054}
\BIBentrySTDinterwordspacing

\bibitem{9149272}
S.~Rezaie, C.~N. Manchón, and E.~de~Carvalho, ``Location- and orientation-aided millimeter wave beam selection using deep learning,'' 2020, pp. 1--6.

\bibitem{9388790}
M.~Wang, Y.~Lin, Q.~Tian, and G.~Si, ``Transfer learning promotes {6G} wireless communications: Recent advances and future challenges,'' \emph{IEEE Transactions on Reliability}, vol.~70, no.~2, pp. 790--807, 2021.

\bibitem{9580346}
H.~Chen, C.~Sun, F.~Jiang, and J.~Jiang, ``Beams selection for mmwave multi-connection based on sub-6{GH}z predicting and parallel transfer learning,'' in \emph{2021 IEEE/CIC International Conference on Communications in China (ICCC)}, 2021, pp. 469--474.

\bibitem{9048929}
M.~Alrabeiah and A.~Alkhateeb, ``Deep learning for {TDD} and {FDD} massive mimo: Mapping channels in space and frequency,'' in \emph{2019 53rd Asilomar Conference on Signals, Systems, and Computers}, 2019, pp. 1465--1470.

\bibitem{9918162}
A.~Ichkov, S.~Häger, P.~Mähönen, and L.~Simić, ``Comparative evaluation of millimeter-wave beamsteering algorithms using outdoor phased antenna array measurements,'' in \emph{2022 19th Annual IEEE International Conference on Sensing, Communication, and Networking (SECON)}, 2022, pp. 497--505.

\bibitem{remcom_wireless_insite}
{Remcom}, ``Wireless insite,'' \url{http://www.remcom.com/wireless-insite}, accessed: 2024-08-01.

\bibitem{9217146}
A.~Ichkov, P.~Mähönen, and L.~Simić, ``Is ray-tracing viable for millimeter-wave networking studies?'' 2020, pp. 1--7.

\end{thebibliography}

\end{document}